\documentclass[doublecol]{epl2} 

\title{Aperiodic quantum oscillations of particle-hole asymmetric Dirac cones}
\shorttitle{Aperiodic quantum oscillations of particle-hole asymmetric Dirac cones} 

\author{E. Tisserond\inst{1}\and J. N. Fuchs\inst{1,2} \and M. O. Goerbig\inst{1} \and P. Auban-Senzier\inst{1} \and C. M\'ezi\`ere\inst{3} \and P. Batail\inst{3} \and Y. Kawasugi\inst{4} \and M. Suda\inst{5} \and H. M. Yamamoto\inst{5} \and R. Kato\inst{4} \and N. Tajima\inst{4,5,6} \and M. Monteverde\inst{1}}
\shortauthor{E. Tisserond \etal}

\institute{                    
  \inst{1} Laboratoire de Physique des Solides, CNRS UMR $8502$ - Universit\'e Paris-Sud, $91405$ Orsay Cedex, France\\
  \inst{2} Laboratoire  de  Physique  Th\'eorique  de  la  Mati\`ere  Condens\'ee,  CNRS  UMR  $7600$ -
Universit\'e  Pierre  et  Marie  Curie,  $4$  Place  Jussieu,  $75252$  Paris  Cedex  $05$,  France\\
 \inst{3} MOLTECH-Anjou, CNRS UMR $6200$ - Universit\'e d'Angers, B\^atiment K, $49045$ Angers Cedex, France\\ 
 \inst{4} RIKEN, Hirosawa $2$-$1$, Wako, Saitama $351$-$0198$, Japan \\
 \inst{5} Institute for Molecular Science, Okazaki, Aichi $444$-$8585$, Japan \\
  \inst{6} Department of Physics, Toho University, Miyama $2$-$2$-$1$, Funabashi, Chiba $274$-$8510$, Japan
}

\abstract{
We report experimental measurements and theoretical analysis of Shubnikov--de Haas (SdH) oscillations in a Dirac cone system: 
the $\alpha$-(BEDT-TTF)$_{2}$I$_{3}$ organic metal under hydrostatic pressure. The measured SdH oscillations reveal anomalies at high magnetic fields $B$ where the $1/B$ oscillations periodicity is lost above $7\un{T}$. We interpret these unusual results within a theoretical model that takes into account intrinsic distortions of the $\alpha$-(BEDT-TTF)$_{2}$I$_{3}$ Dirac cones such as a parabolic particle-hole asymmetric correction. Others possible causes, such as a cone tilting or a Zeeman effect, are carefully ruled out. The observations are consistent among $\alpha$-(BEDT-TTF)$_{2}$I$_{3}$ samples with different Fermi levels.}

\begin{document}

\maketitle


\section{Introduction}

The isolation of graphene in 2004~\cite{graphene1,graphene2} opened a new field of research for condensed matter physicists, called Dirac physics that continues to fascinate today. 
Belonging to the family of the first synthesized quasi $2$D organic conductors, the $\alpha$-(BEDT-TTF)$_{2}$I$_{3}$ ($=\alpha$I$_{3}$) material, 
which consists of an alternation of insulating planes (iodine planes) and conductive planes (BEDT-TTF planes), has been known and studied since the 
1980's~\cite{alpha1,alpha2,alpha3,alpha4,alpha5}. However, a renewed interest in this salt has followed the highlighting of Dirac charge carriers emerging 
under hydrostatic pressure~\cite{alphaPressure}. Indeed, band-structure calculations and magnetotransport experiments have revealed the presence of 
Dirac fermions under high pressure ($P>1.5\un{GPa}$) in $\alpha$I$_{3}$. However, Dirac physics in $\alpha$I$_{3}$ differs from that in graphene by several aspects. 
First of all, the Dirac cones in $\alpha$I$_{3}$ are tilted leading to a renormalization of the cone velocity, which is one order of magnitude smaller than in graphene~\cite{alphaPressure,alphaTiltedCone1,alphaTiltedCone1b,alphaTiltedCone2}, due also to a larger lattice spacing. Then, contrary to the case of graphene, its three-dimensional layered structure renders an
experimental control in $\alpha$I$_{3}$ of the homogenous Fermi level, e.g. by the application of a gate voltage, extremely difficult. Finally, in terms of charge carriers, 
the $\alpha$I$_{3}$ physical properties are more complicated than in graphene due to a coexistence between Dirac and massive fermions in the vicinity of the Fermi level. 
This coexistence of different carrier types, 
theoretically predicted by ab initio band-structure calculations~\cite{alphaCoexistenceTheo}, has recently been verified  experimentally by electronic transport measurements, 
performed in the classical regime~\cite{alphaCoexistenceExp,alphaCoexistenceExp2}.

In this paper, we present magnetotransport measurements of two types of $\alpha$I$_{3}$ crystals under high hydrostatic pressure ($P>1.5\un{GPa}$), 
that is in the presence of Dirac fermions, and in the quantum regime. At low magnetic fields, we observe typical SdH oscillations in the $\alpha$I$_{3}$ 
magnetoresistance, as already reported in the literature~\cite{alphaSdHOscillations}. Beyond this standard behavior of the SdH oscillations, we show that, 
at higher magnetic fields ($B>7\un{T}$), these measured 
quantum oscillations become unusual with a deviation from their $1/B$ periodicity. This means that the usual Landau plot (i.e. the index $\tilde{n}$ of minima in the 
magnetoresistance as a function of $1/B$) is no longer linear in the high field limit. This kind of particular behavior has very recently been seen for surface states of 3D topological insulator samples~\cite{TopolInsulatorsExperience1,TopolInsulatorsExperience2,TopolInsulatorsMcKenzie}. However, here we show that the effect is much stronger in $\alpha$I$_{3}$ than in topological insulators. 

The remainder of the paper is organized as follows. In the first part, we present the experimental setup and the results of the magnetotransport measurements, performed in the quantum regime. In the second part, in order to interpret these particular experimental results, we develop a theoretical model based on the specificity of the $\alpha$I$_{3}$ band structure.  


\section{Experimental transport measurements in the quantum regime}

\begin{figure}
\begin{center}
\includegraphics[scale=0.31]{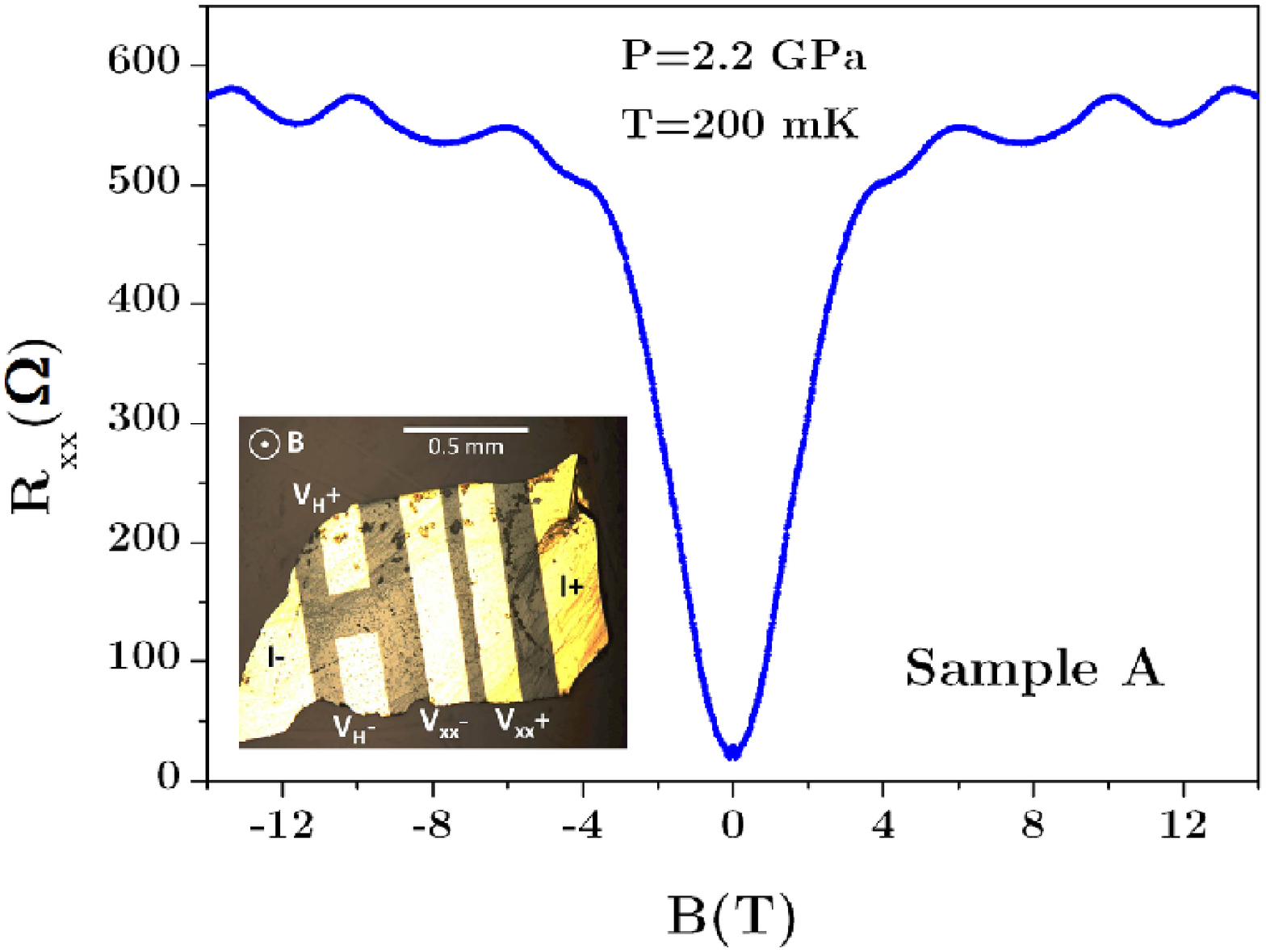}

\includegraphics[scale=0.31]{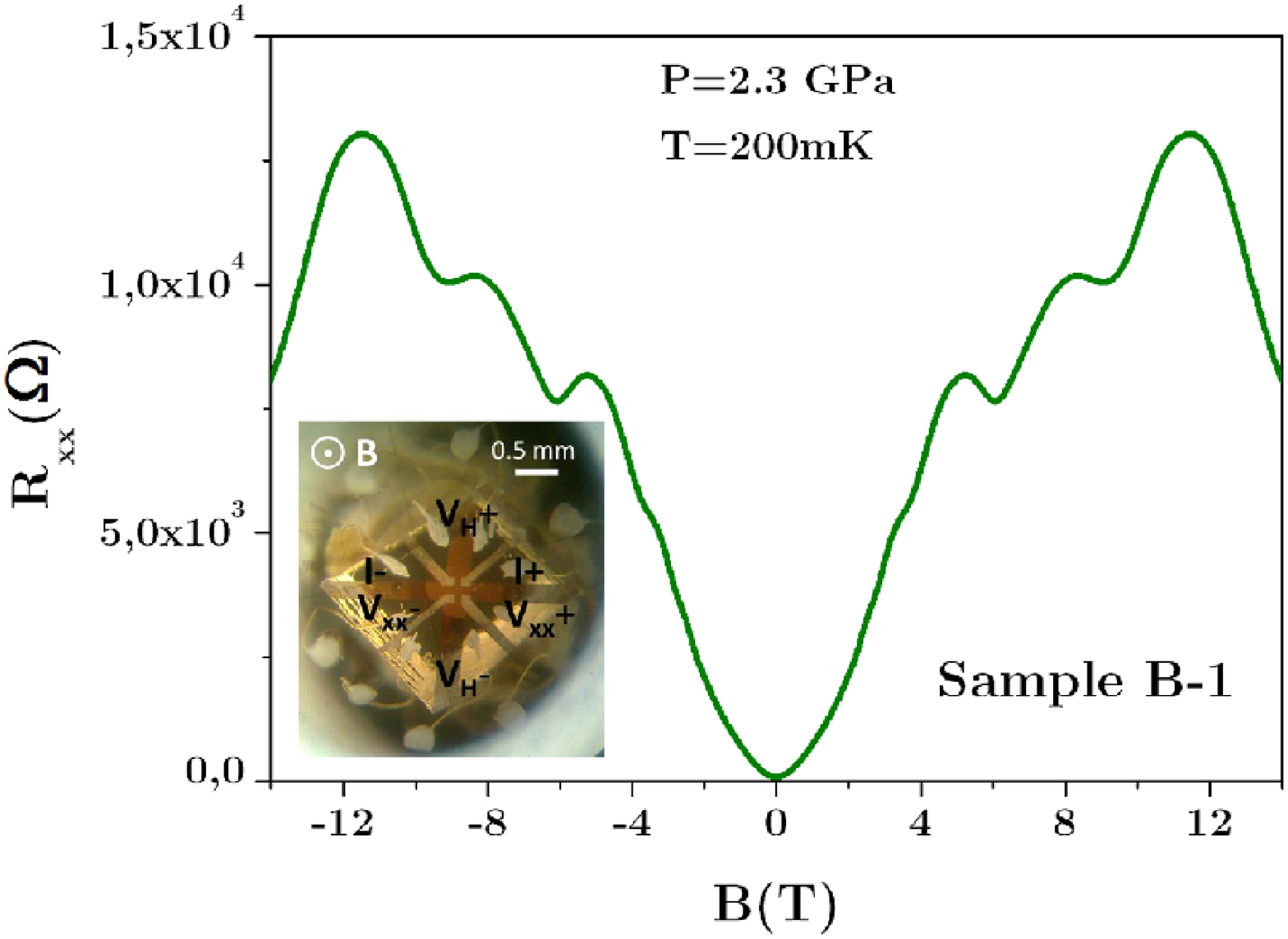}
\end{center}
\vspace{-0.3cm}
\caption{Typical curves of $\alpha$I$_{3}$ magnetoresistance measurements in quantum regime. Quantum oscillations appear on the parabolic classical contribution of the magnetoresistance signal. Top: thick $\alpha$I$_{3}$ crystal. Bottom: thin $\alpha$I$_{3}$ crystal fixed onto a PEN substrate (for details on samples and SdH oscillations, see the supplementary material - part $1$). Insets: Photographs of measured samples.}
\label{fig.1}
\end{figure}

Single crystals of $\alpha$I$_{3}$ were synthesized by electrocrystallisation. In this study, two kinds of sample have been measured: a thick crystal (sample A) and thin crystals fixed onto a polyethylene naphthalate (PEN) substrate (samples B). Their typical size is $1\un{mm^2}$ in the (a-b) plane with a thickness (c direction) of $10\un{\mu m}$ and $100\un{nm}$ for the samples A and B, respectively. The temperature has been controled by means of a dilution fridge. The magnetoresistance (longitudinal signal) and the Hall resistance (transverse signal) have been measured under a magnetic field $B$ oriented along the c direction, perpendicular to the $\alpha$I$_{3}$ conductive planes. The magnetic field was swept between $-14\un{T}$ to $14\un{T}$, at fixed temperature around $200\un{mK}$ and under high hydrostatic pressure between $2.2\un{GPa}$ and $2.6\un{GPa}$. The resistance measurements have been performed simultaneously, using a low-frequency ac lock-in technique, with different types of contacts geometry. For the sample A, we used six gold contacts deposited by Joule evaporation on both sides of the sample and, for the samples B, a Hall cross with eight electrical contacts. The insets of Fig.~\ref{fig.1} show the images of the measured sample A and one measured sample B. The hydrostatic pressure was applied at room temperature in a NiCrAl clamp cell using Daphne $7373$ silicone oil as pressure-transmitting medium and was determined at room temperature by a manganine resistance gauge in the pressure cell, close to the sample. 

\begin{figure}
\begin{center}
\includegraphics[scale=0.31]{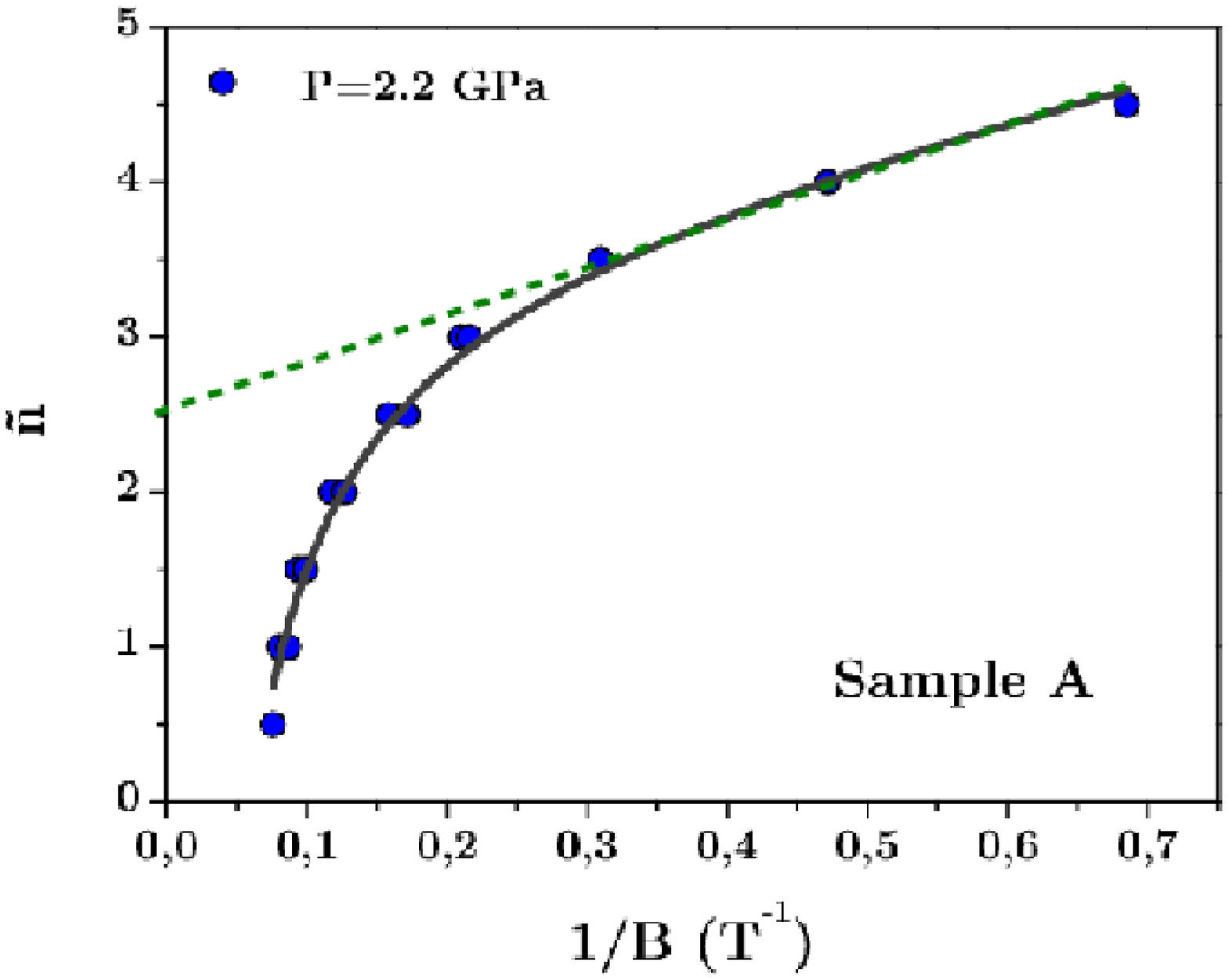}

\includegraphics[scale=0.32]{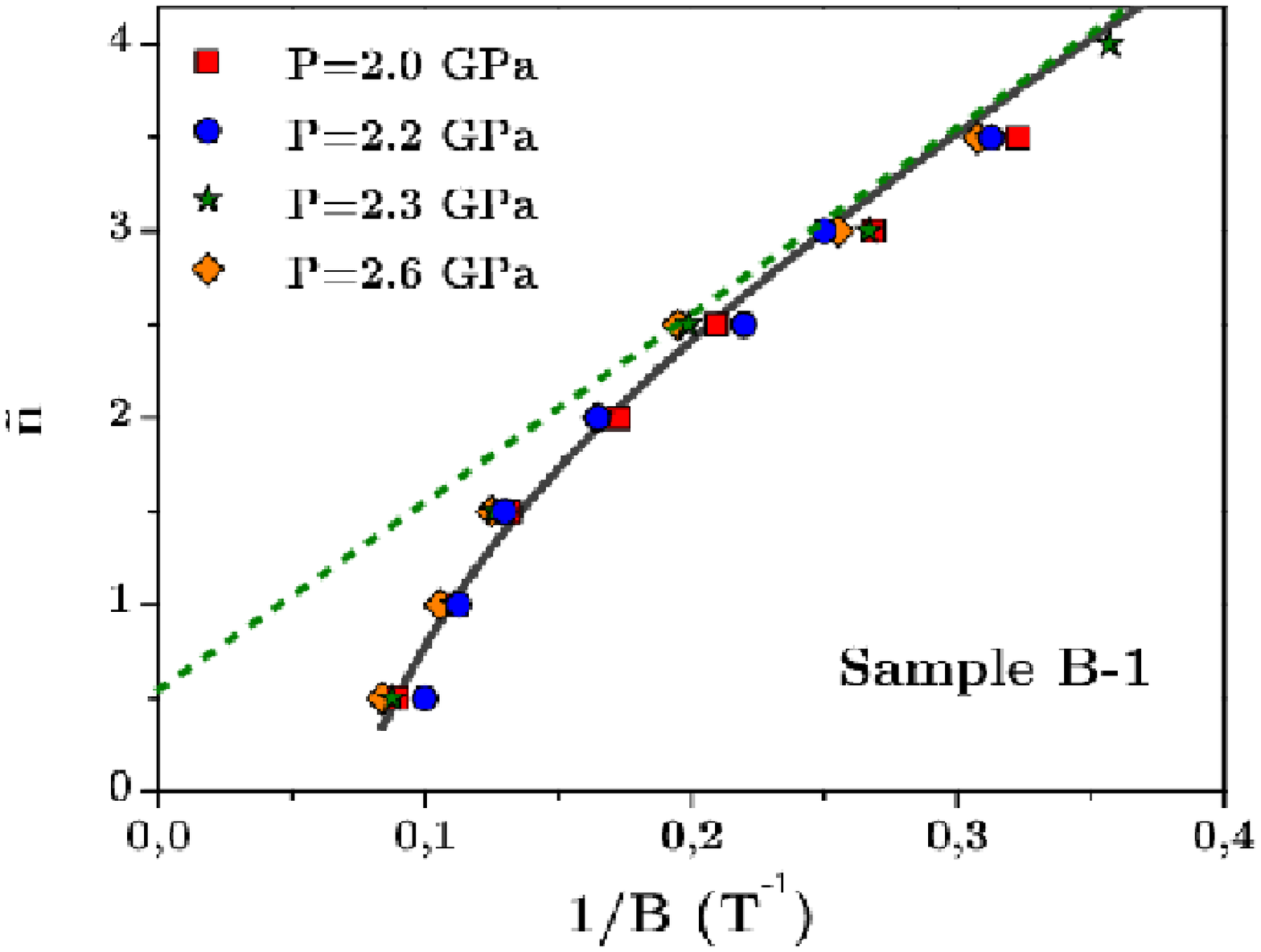}
\end{center}
\vspace{-0.3cm}
\caption{Construction of Landau plots (i.e. $\tilde{n}$ as a function of $1/B$) from the analysis of the measured quantum oscillations in $\alpha$I$_{3}$. At low magnetic fields, the oscillations are $1/B$ periodic: they are SdH oscillations. The determination of their phase offset, connected to the Berry phase, indicates that the Dirac charge carriers are involved in the measured oscillations. At higher magnetic fields, we observe a loss of the $1/B$ periodicity. (dark grey line: theoretical fit of the experimental data points)}
\label{fig.2}
\end{figure}

Typical results of the magnetotransport measurements in the quantum regime are presented in Fig.~\ref{fig.1}. To correct the alignment mismatch of the patterned contacts, the longitudinal resistance has been symmetrized with respect to the magnetic field $B$. We observe clearly the appearance of quantum oscillations on the classical parabolic contribution of the $\alpha$I$_{3}$ magnetoresistance (see Fig.~\ref{fig.1}). Are these measured quantum oscillations $1/B$ periodic, as usual SdH oscillations? Are Dirac carriers at the origin of these oscillations? To answer these two questions, our analysis is based on the study of the magnetoresistance signal and is similar to the one presented in reference~\cite{alphaSdHOscillations}. We plotted the index of the oscillations peaks $\tilde{n}$ (integer for the minima and half-integer for the maxima) as a function of the inverse of their magnetic field $B$ position and we obtained the Landau plots, for both $\alpha$I$_{3}$ crystal types, presented in Fig.~\ref{fig.2}. 
In the range of the studied pressure, the effect of the latter is negligible (see Fig.~\ref{fig.2}, bottom).

At low magnetic fields, the data points are aligned on the Landau plots, and the measured quantum oscillations are thus indeed $1/B$-periodic SdH oscillations. 
Using the usual associated theory~\cite{Shoenberg}, the main harmonic of the oscillating part of the magnetoresistance can be written as 
\begin{equation}
\label{SdHTheory}
\Delta R_{\mathrm{xx}} = A(B) \cos\left(2\pi F/B+\pi +\varphi\right),
\end{equation} 
where $F$ is the magnetic frequency of the oscillations and $\varphi$ is the phase offset associated with the Berry phase, which is $0$ for massive 
fermions and $\pi$ for Dirac 
fermions~\cite{Coleridge,graphene2}. Notice furthermore that $\varphi$ is not necessarily quantized and can take a continuous value between $0$ and 
$\pi$ in the case of more complex band structures~\cite{TopolInsulatorsMcKenzie,MarkPhaseBerryContinuous}. With the previous choice for the index of the 
oscillations peaks, the intercept of the Landau plot $\tilde{n}_{0}$ indicates directly the phase offset 
($\varphi=2\pi\tilde{n}_{0}$). The linear 
extrapolations of our data in the low magnetic field region of the Landau plots give a half-integer $\tilde{n}_{0}$, namely a Berry phase equal to $\pi$, for both 
measured $\alpha$I$_{3}$ crystal types. So, Dirac fermions are involved in the measured quantum oscillations in agreement with the previous reported 
results~\cite{alphaSdHOscillations}. Moreover, the magnetic frequency $F$ -- the slope of the linear fit -- is equal to $2\un{T}$ and to $8.5\un{T}$ for sample 
A and the B-type samples, respectively. Also, this magnetic frequency $F$ is an intrinsic signature of the $2D$ charge carriers density, $n_{2D}$. Indeed, these two 
quantities are related by the following mathematical expression: $F=\left(\phi_{0}/4\right)\times n_{2D}$, where $\phi_{0}=h/e=4.14\times 10^{-15}\un{T m^2}$ is the flux 
quantum and the numerical factor $4$ comes from the four-fold valley and spin degeneracy. By applying this formula, we find $n_{2D}\approx 2\times 10^{11}\un{cm^ {-2}}$ 
for sample A, which corresponds to a value well within previous experimental studies of undoped thick crystals~\cite{alpha3,alphaCoexistenceExp} and doped thin 
crystals~\cite{alphaSdHOscillations}. Meanwhile, for the B type samples, we find $n_{2D}\approx 8\times 10^{11}\un{cm^ {-2}}$ in agreement with the ref.~\cite{alphaSdHOscillations}.

The most salient feature in our magnetotransport data in Fig.~\ref{fig.2} is the deviation from the linear behavior at high magnetic fields ($B> 7$ T), 
where the SdH oscillations are no longer $1/B$-periodic. The theoretical explanation of this deviation is the object of the following section. 


\section{Theoretical interpretation of the measured SdH oscillations}

Several theoretical explanations can be invoked to explain the loss of $1/B$ periodicity at high magnetic fields. First, we could think of identifying a magnetic 
frequency to a given charge carrier type depending on $B$, similarly, for example, to a recent analysis of SdH oscillations measured in some topological insulator 
samples~\cite{SdHVeyrat}. In the $\alpha$I$_{3}$ case, we would have then a first magnetic frequency due to the Dirac carriers and a second one due to the massive carriers. 
We can dismiss this hypothesis because the smooth change of periodicity that we have measured (see Fig.~\ref{fig.2}) is not compatible with the appearance of a second charge 
carriers type involved in the oscillations at a precise magnetic field value. 

Secondly, we could envision a modification of the $1/B$ periodicity due to a cone tilting effect. Previous theoretical works showed that taking into account only the $\alpha$I$_{3}$ Dirac cone tilt gives the same Landau levels structure as in the graphene case with a mere renormalization of the cone velocity 
[$v\rightarrow v\left(1-\beta^ 2\right)^ {3/4}$, where $\beta$ is the dimensionless tilt parameter of typical value in the range of $0.3\ldots 0.8$\footnote{The precise value for the tilt parameter is yet under debate and, to the best of our knowledge, has not been clearly determined.}]~\cite{alphaTiltedCone1,alphaTiltedCone1b,alphaTiltedCone2}. 
This means that the cone tilting alone preserves the $1/B$ periodicity of the quantum oscillations for any field values and does not allow one to explain the experimental results. 

In a third scenario, we could consider a Zeeman effet. In a first approximation, in $\alpha$I$_{3}$, this effet is negligible as the $g$-factor is close to $2$~\cite{gFactor} (see also part $2$ of the supplementary material). 
Moreover, theoretical calculations show that the effect of taking into account this Zeeman contribution leads to a correction of the Landau plot which has the opposite curvature compared to measurements (for more calculation details, see the supplementary material, part $2$). This third hypothesis is therefore not satisfactory either. 

The experimental results described above indicate that it is a behavior proper to the band structure of $\alpha$I$_{3}$ which was probed at high magnetic fields. 
Indeed, in two different $\alpha$I$_{3}$ crystal types, one thick (sample A) and the others thin ones fixed onto a PEN substrate (samples B), the same qualitative 
deviation from the usual SdH theory appears. The origin of this unusual behavior resides then in the intrinsic properties of the $\alpha$I$_{3}$: contrary to the case of 
an ideal linear cone, the Dirac cone is distorted in the $\alpha$I$_{3}$ band structure under pressure~\cite{alphaCoexistenceTheo}. In order to investigate the role of the 
particular band structure of $\alpha$I$_{3}$ on its quantum oscillations and a possible deviation from their $1/B$ periodicity, let us consider the following Hamiltonian 
(for a given valley and spin projection):
\begin{equation}
\label{Hamiltonien}
H=v\left(\Pi_{x}\sigma_{x}+\Pi_{y}\sigma_{y}\right)+\frac{\vec{\Pi}^2}{2m}\sigma_{0},
\end{equation}
where $v\approx 3.5\times 10^{4}\un{m/s}$~\cite{FermiVelocityAlpha} is directly the renormalized cone velocity and $\vec{\Pi}=\vec{p}+e\vec{A}(\vec{r})$ is the gauge-invariant kinetic momentum. For simplicity, we neglect here the explicit role of the Dirac cone tilt and absorb it into the renormalized velocity.
\begin{figure}
\begin{center}
\includegraphics[scale=0.3]{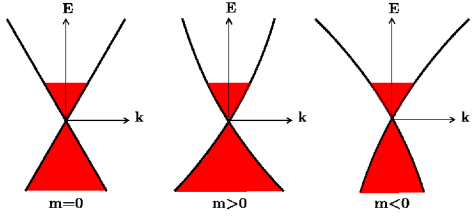}
\end{center}
\caption{We describe theoretically the distorsion of the Dirac cone
in $\alpha$I$_{3}$ by a curvature term with a mass parameter $m$. Left: linear Dirac cone, as in the case of the graphene, middle: distorted Dirac cone ($m>0$), right: distorted Dirac cone ($m<0$).}
\label{fig.3}
\end{figure}
The first term of Eq.~(\ref{Hamiltonien}), which describes the usual Dirac carriers such as in graphene, is completed by a curvature term which formalizes the distortion of 
the $\alpha$I$_{3}$ Dirac cone. The mass parameter $m$ of this curvature term can be positive or negative, depending on the curvature concavity as presented 
in Fig.~\ref{fig.3}. Note that this Hamiltonian breaks particle-hole symmetry, in agreement with previous published 
results~\cite{ParticleHoleSymmetry1,ParticleHoleSymmetry2} [the possibility of a $(\vec{\Pi}^2/2m)\sigma_{z}$ term in our phenomenological model, which 
preserves particle-hole symmetry, has been ruled out because it gives a curvature of the Landau plot with a wrong concavity as compared to the experimental results -- 
see part $2$ of the supplementary material]. Hamiltonian (\ref{Hamiltonien}) was also used to describe graphene in \cite{Sharapov} and is very close to the Rashba model \cite{Rashba}. Here, we only consider the inner -- and neglect the outer -- Fermi surface, as done for surface states of 3D topological insulators and contrary to the Rashba model, see discussion and Fig. 2 in \cite{TopolInsulatorsMcKenzie}.

From Eq.~(\ref{Hamiltonien}), the Landau levels $E_{n}$ can readily be calculated~\cite{MarkPhaseBerryContinuous,Sharapov,Rashba,Wang}
\begin{equation}
E_{n}=\hbar\omega_{m} n\pm \sqrt{\left(\hbar\omega_{v}\right)^2\! n +\left(\frac{\hbar \omega_{m}}{2}\right)^2},
\label{LandauLevelsEnergy}
\end{equation}
where $\omega_{m}=eB/m$, $\omega_{v}= \sqrt{2e v^2 B/\hbar}$ and the Landau level index $n$ is a positive integer such as $n\geq 1$ 
(for $n=0$, $E_{0}=+\frac{\hbar\omega_{m}}{2}$ independently of the valley, see part $2$ of the supplementary material for more details). The positive part of Eq.~(\ref{LandauLevelsEnergy}) corresponds to the conduction-band
contribution and, the negative part, to that of the valence band. We remind that the maxima of both the magnetoconductance and the magnetoresistance correspond to half-filled Landau levels, i.e to peaks in the density of states\cite{Coleridge,Ando}. They appear when $E_{F}=E_{n}$, where $E_{F}$ is the Fermi energy. 
So, we have the following relation between the usual experimental $\tilde{n}$ and theoretical $n$ Landau level index convention: $\tilde{n}=n+1/2$.  

By inverting Eq.~(\ref{LandauLevelsEnergy}), the index $n$ is rewritten as a function of the Fermi energy $E_{F}$ and the magnetic field $B$. In the case of the $\alpha$I$_{3}$, the energy parameter $E_{F}$ is fixed during the crystals growth and can be positive or negative, depending on the natural doping induced by the samples fabrication methods. The Landau level index $n$ is then only a function of the magnetic field $B$: 
\begin{equation}
\begin{array}{lll}
n(B)&=&\left(\frac{m^2v^2}{\hbar e}+\frac{mE_{F}}{\hbar e}\right)/B\\[0.15cm]
&-&\left(\sqrt{\left(\frac{m^2v^2}{\hbar e}\right)^2+2\frac{mE_{F}}{\hbar e}\frac{m^2v^2}{\hbar e}+\left(\frac{B}{2}\right)^2}\right)/B \\[0.15cm]
&=&\left(\frac{\left|m\right|^2v^2}{\hbar e}+s\frac{\left|m\right|\left|E_{F}\right|}{\hbar e}\right)/B\\[0.15cm]
&-&\sqrt{\left(\frac{\left|m\right|^2v^2}{\hbar e}\right)^2+2s\frac{\left|m\right|\left|E_{F}\right|}{\hbar e}\frac{\left|m\right|^2v^2}{\hbar e}+\left(\frac{B}{2}\right)^2}/B,
\end{array}
\label{IndexLandauLevels}
\end{equation}
with $s=\mathrm{sign}(E_{F})\mathrm{sign}(m)$.

Finally, knowing that the parabolic distortion term of the Hamiltonian~(\ref{Hamiltonien}) is a correction compared to the Dirac cone term, we performed an expansion in powers of $B$ (i.e. at small $1/m$) of Eq.~(\ref{IndexLandauLevels}) and we obtained an approximate expression of the Landau level index $n$ as a function of the magnetic field $B$, with two fitting parameters $F$ and $C$:
\begin{equation}
n(B)\approx F\times \frac{1}{B}+ n_0 \times B^0 + C\times B + \mathcal{O}(B^{3}), 
\label{AsymptoticDevelopment}
\end{equation}
with
\begin{equation}
\left\lbrace
\begin{array}{lll}
F
&=&\frac{\left|m\right|^2v^2}{\hbar e}\left(1+s\frac{\left|E_{F}\right|}{\left|m\right|v^2}-\sqrt{1+2s\frac{\left|E_{F}\right|}{\left|m\right|v^2}}\right)\\[0.4cm]
C 
&=&-\frac{\hbar e}{8\left|m\right|^2v^2}\frac{1}{\sqrt{1+2s\frac{\left|E_{F}\right|}{\left|m\right|v^2}}}
\end{array}
\right. .
\label{parametersFit}
\end{equation}
The first term of Eq.~(\ref{AsymptoticDevelopment}) is the magnetic frequency $F$ responsible for the usual SdH theory dependence on $1/B$, while the second
term is the curvature $C$ which represents the deviation from this $1/B$ dependence, and therefore the $1/B$ periodicity. Notice furthermore that the constant, $B$-field independent, 
offset (proportional to $B^0$) is $n_0=0$ here, in agreement with Dirac carriers. 

We applied this phenomenological model to our magnetotransport measurements and the result of the fit is presented in Fig.~\ref{fig.2} for both measured $\alpha$I$_{3}$ 
crystal types. There is a good agreement between the theoretical fit and the experimental data (see Fig.~\ref{fig.2}), moreover the obtained $F$ value is 
in agreement with the low-field linear slope. In Table~\ref{Table} are listed the quantities $n_{2D}$, $T_{F}$ and $m$ deduced from the two fitting parameters. 
The mass parameter $m$, which describes the intrinsic distortion of the $\alpha$I$_{3}$ Dirac cone is well found to be roughly the same (within the fitting uncertainty) 
for both types of $\alpha$I$_{3}$ samples (see Table~\ref{Table}) by choosing $s=+1$ (for details of the fitting, see the supplementary material, part $3$). 
Then, $E_{F}$ and $m$ have the same sign and the nature of the doping (hole or electron doping) is also the same for both measured $\alpha$I$_{3}$ crystal types. 
The Fermi temperature $T_{F}$ in the case of the samples B is higher than that of the sample A, which indicates a more important doping in the B-type samples. This is 
a consequence of the PEN substrate as had been stated elsewhere\cite{alphaSdHOscillations}.     

\begin{table}
\caption{Parameters for both $\alpha$I$_{3}$ sample types in the case of $s=+1$, for which the Fermi energy $E_{F}$ and the mass parameter $m$ have the same sign. We denote as $m_e$ the bare electron mass and we recall that we have taken $v\approx 3.5\times 10^4\un{m/s}$~\cite{FermiVelocityAlpha}.}
\label{Table}
\begin{center}
\begin{tabular}{lcc}
\hline\hline Quantity & Sample A & Samples B\\[0.1cm]\hline
$F \left(\un{T}\right)$ & $2.5\pm 0.5$ & $8.5\pm 0.5$\\[0.1cm]
$C \left(\un{T^{-1}}\right)$ & $-0.22\pm 0.02$ & $-0.17\pm 0.03$\\[0.1cm]
$n_{2D} \left(\un{cm^{-2}}\right)$ & $(2.4\pm0.5)\times10^{11}$ & $(8.3\pm0.5)\times10^{11}$\\[0.1cm]
$T_{F} \left(\un{K}\right)$ & $140\pm44$ & $570\pm130$\\[0.1cm]
$\left|m\right| \left(m_{e}\right)$ & $0.030\pm0.005$ & $0.022\pm0.005$\\\hline\hline
\end{tabular}
\end{center}
\end{table}

\section{Conclusion}
To conclude, we presented $\alpha$I$_{3}$ magnetotransport measurements performed on two different sample types, in the quantum regime and under high hydrostatic pressure. 
We mesured quantum SdH oscillations in the $\alpha$I$_{3}$ magnetoresistance and unveil an unusual behavior under high magnetic fields with a loss of the
characteristic $1/B$ periodicity above $7\un{T}$. 
We show, within a theoretical model that takes into account deviations from the linear shape of the $\alpha$I$_{3}$ Dirac cones, 
that this anomaly can be attributed to a parabolic band correction breaking particle-hole symmetry.  
For both measured $\alpha$I$_{3}$ sample types, there is a good agreement between the experimental data 
and the theoretical fit, which gives reasonable and consistent fit parameters. Indeed, we find a curvature parameter $m$ independent (within the fitting uncertainty) 
from the measured $\alpha$I$_{3}$ sample type with different Fermi levels. The proposed interpretation then provides a suitable background to understand these unusual 
experimental results. We can also note that the distortion of the band structure Dirac cones is at quite low energies which would be a challenge for an independent 
comparison with ab initio band-structure calculations.

\acknowledgments
We acknowledge G. Montambaux, A. Murani, F. Pi\'echon, S. Tchoumakov and L. Veyrat for fruitful discussions. This work was partly supported by Grants-in-Aid for Scientific Research (No.$25287089$, No.$15$H$02108$ and No.$16$H$06346$) from the Ministry of Education, Culture, Sports, Science and Technology, Japan and Nanotechnology Platform Program (Molecule and Material Synthesis) of the Ministry od Education, Culture, Sports, Science and Technology (MEXT), Japan.

\end{document}